\documentclass{optica-article}

\journal{opticajournal} 

\articletype{Research Article}

\usepackage{lineno}

\usepackage{float}
\usepackage{booktabs}
\usepackage{pifont}
\usepackage{multirow}
\usepackage{multicol}
\usepackage{makecell}
\usepackage{graphicx}
\usepackage{mathtools}

\begin{document}

\title{A unified perspective on wavelength selection for molecular composition inference from diffuse spectroscopy}

\author{
Kevin Scibilia,\authormark{1} 
Marcel Rosier,\authormark{1}
Luca Giannoni,\authormark{5}
Pietro Ricci,\authormark{2,3}
Frédéric Lange,\authormark{5}
Florian Kofler,\authormark{6,7,8,9}
Francesco S. Pavone,\authormark{2,3,4}
Ilias Tachtsidis,\authormark{5}
Daniel Rueckert,\authormark{1,10,11}
and Ivan Ezhov\authormark{1}
}

\address{
\authormark{1}Chair for AI in Healthcare and Medicine, Technical University of Munich (TUM) and TUM University Hospital, Germany\\
\authormark{2}Department of Physics and Astronomy, University of Florence, Italy\\
\authormark{3}European Laboratory for Non-Linear Spectroscopy, Italy\\
\authormark{4}National Institute of Optics, National Research Council, Italy\\
\authormark{5}Department of Medical Physics and Biomedical Engineering, University College London, United Kingdom\\
\authormark{6}Hertie Institute for AI in Brain Health, University of Tübingen, Germany\\
\authormark{7}AI for Image-Guided Diagnosis and Therapy, TUM School of Medicine and Health, Technical University of Munich, Germany\\
\authormark{8}Department of Quantitative Biomedicine, University of Zurich, Switzerland\\
\authormark{9}Helmholtz AI, Helmholtz Munich, Germany\\
\authormark{10}Munich Center for Machine Learning (MCML), Germany\\
\authormark{11}Department of Computing, Imperial College London, United Kingdom\\
}

\begin{abstract*} Optical monitoring of living tissue targeting quantification of molecular composition is an active area of research. In the case of broadband spectroscopy, one can attempt to extract molecular, or more precisely, chromophore concentration from a broad-range spectrum of the reflected light.
However, selecting the shortest wavelength range sufficient for quantitative optical monitoring remains an open problem.

Various wavelength optimization methods are scattered throughout the literature, however, there is no unified view to date. Our work's motivation is to construct a wavelength selection framework by unifying existing selection approaches and propose a novel projection-based method that allows for the pre-identification of a wavelength range that is adequate for the final selection. The framework specifically focuses on proposing different methods that quantify match or mismatch between the chosen light-matter interaction model, defined by the chosen endmembers (e.g. molecular chromophores), and the measured intensity from the spectroscopic data. To evaluate the framework, we perform a retrospective analysis on a broadband spectroscopy dataset of piglets during an induced hypoxia-ischemia state. Overall, we show that our novel projection-based method can be used for band selection, and that existing approaches can be used in conjunction to select an optimal minimal wavelength set that satisfies biophysical model constraints. The code is available here: https://github.com/IvanEz/unified-wavelength-selection.

\end{abstract*}




\section{Introduction}

In a spectroscopic application, when light enters biological tissue, the preferred physical viewpoint is within transport theory instead of the Maxwell equations, as biological tissue is non-homogeneous \cite{Cheong1990}. This model assumes that light is transported by individual photons, that are either absorbed or scattered elastically \cite{Wilson1990}. Inclusion of these two effects was empirically determined to be sufficient in clinical practical scenarios \cite{Wilson1990}, instead of using more complex light interaction models. 

Absorption can be modeled in a relatively simple manner via the Beer-Lambert Law. The Beer-Lambert Law (BLL) \cite{Oshina2021, Pichette2016, Kohl2000} states that light absorbed
by a substance is directly proportional to the concentration of the substance:
\[
A\left(\lambda\right)=d\sum_{i}\epsilon_{i}(\lambda)\cdot c_{i}
\]
with $A$ the absorbance, $\epsilon_i$ the extinction coefficient of molecule $i$ with
unit $\left[cm^{-1} \cdot M^{-1}\right]$, $c_{i}$ - the molar concentration,
$d$ the thickness of the medium in cm, and $M$ the molar concentration
in Mol. 
Modification of this model \cite{Delpy1988} allows to account for
multiple scattering \cite{Oshina2021,giannoniphd}, which is commonly present
in optical measurements of biological tissue. This yields the Modified BLL (MBLL) as 
\[
A(t,\lambda)=\sum_{i} \text{PL}(\lambda) \cdot \epsilon_{i}(\lambda)\cdot c_{i} + G(t,\lambda)
\]
where $G(t,\lambda)$ is a geometrical factor quantifying scattering-related
loss of light intensity. PL$(\lambda)$ is the average
optical pathlength  \cite{Pichette2016,Scholkmann2014} travelled by photons in tissue, from source to
detector, defined as 
\[
\text{PL}(\lambda)=\text{DPF}(\lambda)\cdot d
\]
where DPF$(\lambda)$ is the differential pathlength factor accounting for wavelength-dependent increments in the distance travelled by photons due to multiple scattering. 

We now emphasize a few key assumptions that allow us to simplify the
problem by inferring changes in molecular concentrations:
\begin{enumerate}
	\item The incident light intensity $I_{0}$ is constant in time \cite{giannoniphd}.
	\item Changes over time in scattering properties $G(t,\lambda)$ are negligible
	compared to changes in absorption \cite{Oshina2021,Kocsis2006,giannoniphd}.
	\item Changes in attenuation are small over time and homogeneous over space
	\cite{Oshina2021,Kocsis2006}.
\end{enumerate}
By assuming a scattering loss $G(t,\lambda)$ that is constant in
time (or changes at least being small compared to absorption \cite{Giannoni2018, giannoni2026hyperspectral}),
one can significantly simplify the problem of inferring molecular
concentrations \cite{Kocsis2006}. Selection of a baseline attenuation
$A(t_{0},\lambda)$ at timepoint $t_{0}$ and comparison with the
attenuation at timepoint $t_{i}$ by subtraction yields the so-called
differential MBLL (dMBLL):
\[
\Delta A\left(t_{1},\lambda\right)=A(t_{1},\lambda)-A(t_{0},\lambda)=\log\left(\frac{I_{0}}{I(t_{1},\lambda)}\right)-\log\left(\frac{I_{0}}{I(t_{0},\lambda)}\right)=-\log\left(\frac{I(t_{1},\lambda)}{I(t_{0},\lambda)}\right)
\]
This corresponds to
\[
\Delta A\left(t_{1},\lambda\right)=\left(\sum_{i} \text{PL}(\lambda) \cdot \epsilon_{i}(\lambda) \cdot c_{i}(t_{1})+G(\lambda)\right)-\left(\sum_{i} \text{PL}(\lambda) \cdot \epsilon_{i}(\lambda)\cdot c_{i}(t_{0})+G(\lambda)\right)=
\]
\[
=\sum_{i} \text{PL}(\lambda) \cdot \epsilon_{i}(\lambda)\cdot\Delta c_{i}
\]
Using this formulation, changes in molecular concentrations can be
inferred for some timepoint $t_{j}$, with respect to $t_{0}$. Assume
that we have measured attenuations across $n$ wavelengths, and that
the pathlength $\text{PL}(\lambda)$ and absorption properties $\epsilon_{j}(\lambda)$
of the $m$ relevant chromophores are known for the chosen wavelengths.
A linear system $\Delta A=M\cdot\Delta c$ can be constructed, where
\begin{equation}
%
 	\Delta A ={-\log\left(\frac{I(t_{j},\lambda_{i})}{I(t_{0},\lambda_{i})}\right)}\in\mathbb{R}^{n\times1}\label{eq:b}
\end{equation}
such that
\begin{equation}
	\Delta A=\begin{bmatrix}\epsilon_{1,1} \cdot \text{PL}(\lambda_1)  & \epsilon_{1,2} \cdot \text{PL}(\lambda_1) & \dots & \epsilon_{1,m} \cdot \text{PL}(\lambda_1)\\
		\epsilon_{2,1} \cdot \text{PL}(\lambda_2) & \epsilon_{2,2} \cdot \text{PL}(\lambda_2) & \dots & \epsilon_{2,m} \cdot \text{PL}(\lambda_2)\\
		\vdots & \vdots & \ddots & \vdots\\
		\epsilon_{n,1} \cdot \text{PL}(\lambda_n) & \epsilon_{n,2} \cdot \text{PL}(\lambda_n) & \dots & \epsilon_{n,m} \cdot \text{PL}(\lambda_n)
	\end{bmatrix}\text{\ensuremath{\cdot}}\begin{bmatrix}\Delta c_{1}\\
		\Delta c_{2}\\
		\vdots\\
		\Delta c_{m}
	\end{bmatrix}\label{eq:b_eq_M_c}
\end{equation}
with $\epsilon_{i,j}$ the absorption coefficient at wavelength $i$ of
chromophore $j$, and $M\in\mathbb{R}^{n\times m}$$,c\in\mathbb{R}^{m\times1}$.
The changes in concentrations $\Delta c$ can now be estimated by
minimizing the least squares problem $\Vert M \Delta c- \Delta A\Vert_{2}^{2}$.
Usually there are more measured wavelengths $n$ than relevant chromophores
$m$, yielding an overdetermined system. An efficient method in solving
such a least squares problem for overdetermined systems is through
the use of the Moore-Penrose pseudoinverse $M^{\dagger}$ that can
be computed through singular value decomposition \cite{Penrose1956,giannoniphd}.
The changes in concentrations can then be found via a matrix-vector
multiplication,
\begin{equation}
	\Delta c=M^{\dagger}\Delta A
	\label{eq:cPI}
\end{equation}

\section{Methods}
In section \ref{sec:current-methods}, we provide an overview of algorithms that can be used for wavelength selection. Then in section \ref{sec:a-unified-perspective}, we show how given a dataset of acquired spectroscopy measurements (from tissue of unknown composition) and absorption
spectra (of possible tissue chromophores), one can identify a reduced relevant set of wavelengths together with chromophores.

\subsection{Overview of Current Methods}\label{sec:current-methods}
Firstly, the task of selecting wavelengths can be seen as a feature selection problem \cite{Ayala2022, Jain1997}, independent of the down-stream application (e.g. classification or inference of molecular concentrations). The problem is defined as follows: given a set of candidate features $Y$ (in our case, all wavelengths used by a spectroscopy device) with $|Y|=n$, we are interested in finding a subset (of wavelengths) $X \subseteq Y$ of size $d$ (i.e. $|X|=d$) where 
\begin{equation}
X = \max_{Z \subseteq Y, |Z| = d} J(Z) \text{  or  } X = \min_{Z \subseteq Y, |Z| = d} J(Z)
\end{equation}
The function $J$ is the feature selection criterion to be maximized (or interchangeably minimized), which will depend on the exact spectroscopic task. There are a total of $\binom{n}{d}$ possibilities when searching for a wavelength subset of size $d$. For most applications, a brute-force search will oftentimes be unfeasible. As an example, imagine to measure the wavelength range from 780$nm$-900$nm$ with 1$nm$ spacing, similarly to \cite{Kaynezhad2019}, and that $J(Z)$ of a candidate wavelength combination $Z$ can be evaluated in $0.1ms = 10^{-4} s$. Finding the optimal set of 15 wavelengths would require $\binom{120}{15} \cdot 10^{-4}s \approx 10^{18} \cdot 10^{-4}s \approx 10^6$ years. While one could argue to use parallel computing and a more coarse wavelength candidate spacing to reduce $n$, the range is however oftentimes larger than in the mentioned example, e.g. in hyperspectral imaging ranging from 400nm to 1000nm. To solve this problem more efficiently, there are three different categories of algorithms that can be used \cite{Jain1997}: sequential methods, stochastic approaches, and optimal methods. 

Sequential methods are deterministic algorithms that iteratively add or remove a wavelength from a candidate set. They either start with an empty set and add wavelengths (forward methods), or start with a full set and delete wavelengths (backward methods). The method of iteratively adding one wavelength (the most relevant one maximising $J$) is called "Sequential Forward Selection" (SFS), and the method of starting with all wavelengths and iteratively removing one (the least relevant one, i.e. the one reducing $J$ the least) is called "Sequential Backward Selection" (SBS). More complex algorithmic variants exist \cite{Jain1997}, such as the Sequential Forward Floating Selection method.

Stochastic approaches entail the use of, e.g. Genetic Algorithms (GA) \cite{Katoch2021}, to find the optimal wavelengths. Using stochastic methods seems to be the most widely used method for selecting wavelengths in the literature. The authors in \cite{Arifler2015} and \cite{Chauvin2022} used a genetic algorithm and simulated annealing \cite{Kirkpatrick1983} method to find optimal wavelengths when using the modified BLL and an extended version thereof, respectively. The work \cite{Martinez2019} also used different stochastic algorithms to find the optimal wavelengths in a hyperspectral imaging scenario of brain tumors, including a Genetic Algorithm, Particle Swarm Optimization \cite{Perez2007}, and Ant Colony Optimization \cite{Sharma2018}. Another possibility is using the Differential Evolution algorithm, as shown previously in \cite{An2015, Storn1997}.


The above algorithms can be used for any sensible choice of the target function $J$ to be optimized for. An alternative viewpoint on feature selection, especially with regards to multispectral and hyperspectral imaging, can also be found in the literature \cite{Ayala2022,Guyon2003}. According to it, we can differentiate between 'wrapper' and 'filter' methods. Wrapper methods have the goal of maximizing or minimizing some target performance metric, which is equivalent to the target function $J$ formulation we have seen previously. Filters determine the features as a pre-processing step and are therefore unaware of any task $J$ that might be imposed. 

Having seen different definitions for feature and wavelength selection, we will now devote ourselves to looking at more concrete examples.
For hyperspectral imaging, we divide the methods into three distinct categories: methods that have the goal of preserving inferred molecular changes, methods that focus on model-orthogonality and approaches focusing on data-orthogonality. These are explained in greater detail in the following sections.
\subsubsection{Preservation of Inferred Molecular Changes}\label{sec:preservation-concentrations}

One intuitive and widely used approach to select wavelengths is to base the selection process on the preservation of inferred molecular concentrations from the complete wavelength range \cite{Arifler2015, Chauvin2022, Ayala2022}. The most relevant example from the literature for this work is \cite{Arifler2015}, as it selects wavelengths within the context of inferring concentrations from broadband near-infrared spectroscopy (bNIRS) of piglets experiencing hypoxia-ischemia (HI) via the MBLL formulation. The target function $J$ to minimize was chosen as 
\begin{equation}
J(Z) = \text{NRMSE}\left(\Delta c-\Delta c_{Z}\right) = \text{NRMSE}\left(M^{\dagger}\Delta A-M_{Z}^{\dagger}\Delta A_{Z}\right).
\end{equation}
For some subset $Z$ of wavelengths, the function $J$ computes the normalized root mean squared error (NRMSE) between the inferred concentrations $\Delta c$ obtained by using the MBLL with all available wavelengths, and the concentrations $\Delta c_Z$ obtained only by using the wavelenghts in $Z$. The concentrations can be inferred as we previously described in equation \ref{eq:cPI}, noting that $M_Z^{\dagger}$ and $\Delta A_{Z}$ are the absorption matrix and the attenuation vector only with the rows that correspond to the wavelengths in $Z$. 
As the concentrations are inferred for multiple timepoints, the above definition is extended to 
\begin{equation}
J(Z) = \text{NRMSE}\left[ \sum_{t_{i}=1}^{t_{n}} \left( \Delta c(t_i)-\Delta c_{Z}(t_i) \right) \right]
\end{equation}
for all timepoints $t_i=1,\dots,t_n.$ A more detailed overview of different selected wavelengths specifically for the bNIRS HI piglet scenario is provided in \cite{Bale2016}. 

\subsubsection{Model Orthogonality-Based Approaches}\label{sec:model-orthogonality}

An alternative approach is to choose a function $J$ that is directly connected to the physical model being used and maximise the orthogonality within model features. A more concrete example with the BLL (and modifications thereof) would be e.g. the maximization of orthogonality of the absorption features or the maximization of posedness for the inverse problem \cite{Luke2013,Marois2018,Eames2008,Sun2022,Corlu2003,Corlu2005,Correia2010,Arridge1998,Mazhar2010}. 

In \cite{Marois2018,Corlu2003,Luke2013,Mazhar2010}, the wavelengths are chosen based on a criterion $J$ using the singular values of the absorption matrix $M$. The magnitude of singular values is known to describe orthogonality \cite{Marois2018}, and therefore, one can consider the preservation of the magnitude of the singular values as a possible wavelength selection criterion.
The motivation of such approaches is that maximizing the orthogonality of the spectra in $M'$ should remove any wavelengths that contain redundant spectral information. 
The typical choice of $J$ is the maximization of the minimum singular value \cite{Luke2013} or of the condition number \cite{Corlu2003,Luke2013,Mazhar2010}, as described by \cite{Marois2018}. 
The authors in \cite{Marois2018} however argue that a better criterion is the product of singular values, as it uses all singular values instead of only using one or two as in the previously mentioned choices for $J$. 
Finally, the algorithm used to maximize $J$ in \cite{Luke2013,Marois2018,Luke2014} resembles the classical SBS algorithm introduced in section \ref{sec:current-methods}. 

For our study, we apply the SBS algorithm to the absorption matrix having the product of singular values as the optimization objective.


\subsubsection{Data Orthogonality-Based Approaches}\label{sec:data-orthogonality}
We have discussed different wrapper methods and possibilities of choosing $J$; however, as mentioned above, there is also a different class of the so-called filter methods \cite{Chang2006,Nouri2016,Peng2005,Ding2005}. These methods are often based on choosing wavelengths that maximally preserve the information of the measured (e.g. multispectral or hyperspectral) data, independent of the underlying physical model. One might argue that such methods are not of interest for biophysical-model-based inference of changes in concentrations, as one would discard e.g. absorption features of the model. However, we believe that aligning model-based and model-independent wavelength selection approaches can yield valuable insights, such as evaluating the suitability of a particular model to the given spectral data, or identifying the presence of unmodeled physical phenomena and hidden endmembers (e.g. chromophores).

In this work, we resort to the SBS algorithm to optimise for data orthogonality. In contrast to the model orthogonality, the matrix for which we compute the singular values is composed of measured attenuation spectra $\Delta A$. 
Essentially, every column of the matrix contains a spectrum from the spectroscopy dataset. 
The algorithm will select bands at which the spectra across the dataset differ most, explaining the largest variability in the dataset. We believe the selected bands will likely be the most useful for the downstream task, e.g. tissue status classification.

\subsection{A Unified Perspective}\label{sec:a-unified-perspective}
\begin{figure}[h]
	\begin{centering}
		\includegraphics[width=\textwidth]{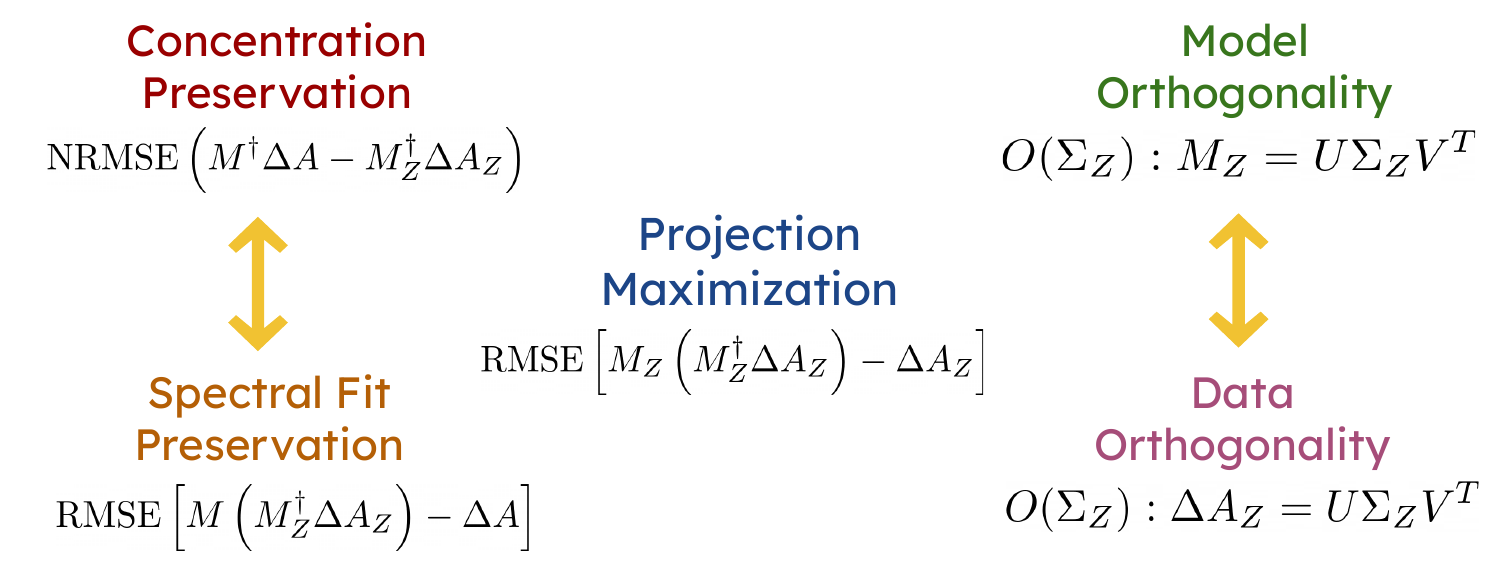}
		\par\end{centering}
	\caption{Overview of the different presented wavelength selection methods, with notation as introduced in the text. Concentration Preservation has a similar counterpart that we denote "Spectral Fit Preservation". Model Orthogonality also has a respective counterpart, which is denoted as Data Orthogonality, while "Projection Maximization" is not directly comparable with the other methods.}
	\label{fig:wavelength-selection}
\end{figure}

We have already formalized and discussed the different existing paradigms of preserving model-inferred concentrations, model and data orthogonality. We now extend those paradigms by including two additional methods of selecting wavelengths and
visualize all approaches in Fig. \ref{fig:wavelength-selection}.

\subsubsection{Spectral Fit Preservation} 
The number of wavelengths can be minimized to preserve the spectral fit obtained by using all wavelengths, which we coin here as "Spectral Fit Preservation". This method focuses on maximizing the model fit over the whole measured range with the reduced subset $Z$, similarly as mentioned in \cite{Bahl2023}. The optimization condition in this case consists of first computing the inferred concentrations with the reduced subset $\Delta c_Z = M_Z^{\dagger} \Delta A_Z$, then comparing with the inferred attenuation obtained over the full measured wavelength spectrum with concentrations $\Delta \tilde{A} = M \Delta c_Z$, and finally computing the error w.r.t. the measured attenuation over the full range: $\text{RMSE} \left[ \Delta \tilde{A} - \Delta A \right]$. 

It might be tempting to assume that the methods preserving concentrations or spectral fit are equivalent, as we would expect a successful preservation of the inferred concentrations to also lead to high spectral fit with respect to the fully measured attenuation over all bands. However, note that the underlying mathematical operation differs. If we rewrite $M^{\dagger} \Delta A - M_Z^{\dagger} \Delta A_Z$ with typical naming conventions for better readability, i.e. assuming 
\begin{equation}
A \coloneqq M^{\dagger}, x \coloneqq \Delta A, c \coloneqq Ax, A_Z \coloneqq M_Z^{\dagger}, x_Z \coloneqq \Delta A_Z, c_Z \coloneqq A_Z x_Z,    
\end{equation} 
we observe that the operation of preserving concentrations corresponds to minimizing $Ax - c_Z$. The spectral fit preservation method opposingly corresponds to the minimization of $A^{\dagger} c_Z - x$. The concentrations $c_Z$ inferred through the reduced wavelength set are simply subtracted from another vector in the former case, whereas being transformed linearly in the latter case. We argue that it is therefore an alternative method of finding wavelengths, differing from the classical method of preserving chromophores concentrations, that needs to be evaluated separately.

\subsubsection{Projection Maximization} We finally introduce a wavelength selection algorithm that we have coined "Projection Maximization". We noted before that one desirable property of wavelength selection algorithms might be to fit the measured signal "well". We have come closest to this objective by preserving spectral fit, minimizing $\text{RMSE} \left[ M \left( M_Z^{\dagger} \Delta A_Z  \right) - \Delta A \right]$. However, the underlying assumption of the full measured attenuation $\Delta A$ being the ground truth to be achieved is not necessarily correct in all cases. For many applications, the underlying model $M$ is yet to be discovered \cite{Ezhov2023}, and certain bands might have information from chromophores or other effects that are not described by $M$ (i.e. scattering). The projection maximization method does not rely on $M$ and $\Delta A$ being the ground truth, but instead uses the already reduced versions $M_Z$ and $\Delta A_Z$. As such, the spectral fit preservation method $\text{RMSE} \left[ M \left( M_Z^{\dagger} \Delta A_Z  \right) - \Delta A \right]$ can be rewritten to use the reduced versions of $M_Z$ and $\Delta A_Z$ to obtain $\text{RMSE} \left[ M_Z \left( M_Z^{\dagger} \Delta A_Z  \right) - \Delta A_Z \right]$. 

It may seem counterintuitive to use $M_Z$ and $\Delta A_Z$ as ground truth, since one could assume that $M_Z (M_Z^{\dagger} \Delta A_Z)$ is always equal to $\Delta A_Z$. However, the inferred concentrations $\Delta c_Z = M_Z^{\dagger} \Delta A_Z$ might not necessarily match the real concentrations from which the signal originated, particularly when the model insufficiently describes the measured attenuation. The model-inferred attenuation $\Delta \hat{A} = M_Z (M_Z^{\dagger} \Delta A_Z)$ will then significantly differ from the real measured attenuation $\Delta A$. We therefore think that removing wavelengths, which the model fails at describing well, is an alternative reasonable approach for wavelength selection. Given a spectroscopy dataset, it can be used during the exploration of different models $M$, as it gives insights into which model describes which bands well. We have coined the method as "Projection Maximization", since the operation $P = M_Z M_Z^{\dagger}$ follows the definition of a vector-space projection, and intuitively projects the measured attenuation into the space of model-based attenuations (i.e. attenuations that can be obtained through the model). The described projection maximization is effectively an optimization of the objective function $J$ that can be achieved with the different approaches we have previously seen in section \ref{sec:current-methods}. For consistency, we optimize the function with the SBS algorithm in this work
\footnote{Note that while the use of such projections in the context of the BLL or during wavelength optimization has been used in previous different settings, see e.g. \cite{Eames2008}, its use both for wavelength optimization and model exploration is novel.}. 


\section{Results}
\subsection{Evaluation on Piglet Broadband NIRS Dataset \& Overview}
In the following, we will evaluate the methods described above using broadband NIRS (bNIRS) spectra of 27 piglets' during an induced hypoxia-ischemia state \cite{Kaynezhad2019}, which has been described in \cite{ezhov2024shallow}. We use the first 1000 measured timepoints, which suffice to observe the induced HI and recovery state. The chromophores to perform model fitting were chosen as in \cite{Kaynezhad2019}, namely using oxyhaemoglobin, deoxyhaemoglobin and differential cytochrome-c-oxidase (diffCCO).

This dataset is particularly suited to benchmark wavelength selection methods of BLL-based methods in biomedical settings, as previous studies have identified specific wavelengths to infer chromophore concentrations while maintaining low estimation errors \cite{Arifler2015}. 

We provide an overview of the presented methods in Table \ref{tab:method_results}, detailing the different requirements they have regarding their input. We highlight in particular our proposed integration with the other wavelength selection methods, as we believe they can complement each other based on the results we discuss below. We will further present the minimal set of wavelengths found across the different tested methods, as summarized in Table \ref{tab:found-wl-combined}. 

\newcommand{\cmark}{\ding{51}}%
\newcommand{\xmark}{\ding{55}}%
\newcommand{\arrr}{\ding{72}}%

\begin{table}[h]
\centering
\footnotesize
\setlength{\tabcolsep}{6pt}
\renewcommand{\arraystretch}{1.15}
\resizebox{\textwidth}{!}{
\begin{tabular}{
>{\centering\arraybackslash}p{3cm}
>{\centering\arraybackslash}c
>{\centering\arraybackslash}c
>{\centering\arraybackslash}c
>{\centering\arraybackslash}c
>{\centering\arraybackslash}p{4.0cm}
}
\toprule

\textbf{Method} &
\makecell[c]{\textbf{Requires}\\\textbf{Model?}} &
\makecell[c]{\textbf{Requires}\\\textbf{Data?}} &
\makecell[c]{\textbf{Use for Band}\\\textbf{Selection?}} &
\makecell[c]{\textbf{Use for Minimal}\\\textbf{Wavelength Selection?}} &
\makecell[c]{\textbf{Proposed}\\\textbf{Integration}} \\

\midrule

Concentration Preservation &
\cmark & \cmark & \xmark & \cmark &
Compare selected wavelengths across methods. \\

Spectral Fit Preservation &
\cmark & \cmark & \xmark & \cmark &
Compare selected wavelengths across methods. \\

Model Orthogonality &
\cmark & \xmark & \xmark & \cmark &
Compare with preservation methods to ensure consistency with measured data. \\

Data Orthogonality &
\xmark & \cmark & \xmark & \xmark &
Avoid when molecular concentrations differ in scales.  \\

Projection Maximization &
\cmark & \cmark & \cmark & \xmark &
Potentially identifies relevant wavelength bands based on selected model candidate. \\

\bottomrule
\end{tabular}
}
\caption{Summary of the presented wavelength selection methods. Each method requires as input certain model assumptions and / or spectral data, and can subsequently infer a relevant band or minimal set of wavelengths. We finally summarize the purpose for each of the selected method. Legend: \cmark: Yes. \xmark: No.}
\label{tab:method_results}
\end{table}

\newcommand{\wln}[1]{\makebox[2.35em][c]{\strut #1}}
\begin{table}[h]
	\centering
	\footnotesize
	\setlength{\tabcolsep}{4pt}
	\renewcommand{\arraystretch}{1.15}
	\resizebox{\textwidth}{!}{
	\begin{tabular}{
		>{\centering\arraybackslash}p{0.8cm}
		>{\centering\arraybackslash}p{3.6cm}
		>{\centering\arraybackslash}p{1.6cm}
		*{8}{>{\centering\arraybackslash}m{0.95cm}}
	}
		\toprule
		\makecell[c]{\textbf{Card.}\\\textbf{$d$}} &
		\makecell[c]{\textbf{Selection}\\\textbf{Criterion}} &
		\makecell[c]{\textbf{Selection}\\\textbf{Algorithm}} &
		\multicolumn{8}{c}{\textbf{Found Wavelengths [nm]}} \\
		\cmidrule(lr){4-11}
		& & &
		\textbf{1} & \textbf{2} & \textbf{3} & \textbf{4} &
		\textbf{5} & \textbf{6} & \textbf{7} & \textbf{8} \\
		\midrule
		\multirow{4}{*}{\makecell[c]{$3$}} &
		Concentration Preservation &
		Brute-force &
		\wln{780.5} & \wln{829.9} & \wln{898.3} & \wln{--} & \wln{--} & \wln{--} & \wln{--} & \wln{--} \\
		&
		Spectral Fit Preservation &
		Brute-force &
		\wln{783.9} & \wln{829.9} & \wln{895.1} & \wln{--} & \wln{--} & \wln{--} & \wln{--} & \wln{--} \\
		&
		Model Orthogonality &
		SBS &
		\wln{780.5} & \wln{824.7} & \wln{899.6} & \wln{--} & \wln{--} & \wln{--} & \wln{--} & \wln{--} \\
		&
		Data Orthogonality &
		SBS &
		\wln{781.8} & \wln{810.9} & \wln{898.9} & \wln{--} & \wln{--} & \wln{--} & \wln{--} & \wln{--} \\
		\midrule
		\multirow{4}{*}{\makecell[c]{$4$}} &
		Concentration Preservation &
		Brute-force &
		\wln{780.5} & \wln{813.5} & \wln{839.7} & \wln{895.1} & \wln{--} & \wln{--} & \wln{--} & \wln{--} \\
		&
		Spectral Fit Preservation &
		Brute-force &
		\wln{780.5} & \wln{803.6} & \wln{852.8} & \wln{895.1} & \wln{--} & \wln{--} & \wln{--} & \wln{--} \\
		&
		Model Orthogonality &
		SBS &
		\wln{780.5} & \wln{824.7} & \wln{825.3} & \wln{899.6} & \wln{--} & \wln{--} & \wln{--} & \wln{--} \\
		&
		Data Orthogonality &
		SBS &
		\wln{781.8} & \wln{783.2} & \wln{810.9} & \wln{898.9} & \wln{--} & \wln{--} & \wln{--} & \wln{--} \\
		\midrule
		\multirow{4}{*}{\makecell[c]{$5$}} &
		Concentration Preservation &
		Brute-force &
		\wln{780.5} & \wln{813.5} & \wln{833.2} & \wln{891.9} & \wln{898.3} & \wln{--} & \wln{--} & \wln{--} \\
		&
		Spectral Fit Preservation &
		Brute-force &
		\wln{780.5} & \wln{803.6} & \wln{833.2} & \wln{875.6} & \wln{898.3} & \wln{--} & \wln{--} & \wln{--} \\
		&
		Model Orthogonality &
		SBS &
		\wln{780.5} & \wln{824.7} & \wln{825.3} & \wln{898.9} & \wln{899.6} & \wln{--} & \wln{--} & \wln{--} \\
		&
		Data Orthogonality &
		SBS &
		\wln{781.8} & \wln{783.2} & \wln{810.9} & \wln{814.8} & \wln{898.9} & \wln{--} & \wln{--} & \wln{--} \\
		\midrule
		\multirow{4}{*}{\makecell[c]{$8$}} &
		Concentration Preservation &
		Brute-force &
		\wln{780.5} & \wln{787.1} & \wln{813.5} & \wln{816.8} & \wln{836.5} & \wln{839.7} & \wln{895.1} & \wln{898.3} \\
		&
		Spectral Fit Preservation &
		Brute-force &
		\wln{780.5} & \wln{787.1} & \wln{803.6} & \wln{820.1} & \wln{839.7} & \wln{869.1} & \wln{885.4} & \wln{891.9} \\
		&
		Model Orthogonality &
		SBS &
		\wln{780.5} & \wln{781.2} & \wln{824.7} & \wln{825.3} & \wln{825.9} & \wln{898.3} & \wln{898.9} & \wln{899.6} \\
		&
		Data Orthogonality &
		SBS &
		\wln{781.8} & \wln{783.2} & \wln{810.9} & \wln{814.8} & \wln{891.9} & \wln{897.0} & \wln{898.9} & \wln{899.6} \\
		\bottomrule
	\end{tabular}
	}
	\caption{Found wavelengths across cardinalities $d \in \{3,4,5,8\}$ for four selection criteria in the fine-grained $[780\,\mathrm{nm},900\,\mathrm{nm}]$ scenario. Concentration and spectral fit preservation were optimized via brute-force search; model and data orthogonality were optimized via sequential backward selection (SBS). Wavelength indices refer to the ordered selected set for each cardinality.}
	\label{tab:found-wl-combined}
\end{table}

\subsection{Concentration \& Spectral Fit Preserving Methods}\label{sec:concentration-spectral-fit-preserving}

\begin{figure}[H]
	\includegraphics[width=0.5\textwidth]{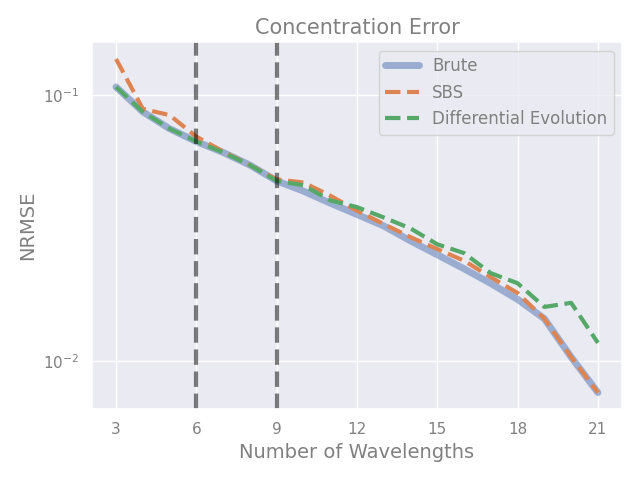}
	\includegraphics[width=0.5\textwidth]{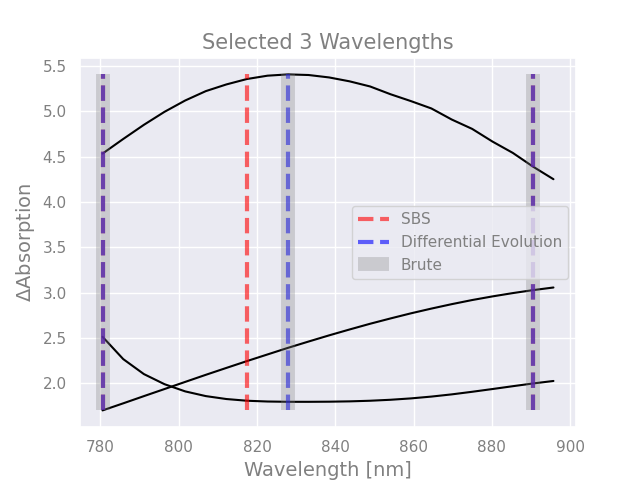}
	\caption{Performance of preserving concentrations selection methods across all possible wavelength combinations, with an example of selected wavelengths. Left: NRMSE development across different cardinalities $d$ of the wavelength set, with corresponding NRMSE with three different algorithms: brute-force search, the SBS algorithm, and Differential Evolution. Right: Absorption of the three main chromophores (oxyhaemoglobin, deoxyhaemoglobin, and diffCCO, colored in black), compared to the wavelengths selected for the case $d=3$ from the three algorithms. The take-home message here is that beyond the [6,9] range of selected wavelengths, different algorithms result in different selections: below 6 wavelengths, brute-force search and Differential Evolution agree, while above 9 - brute-force search and SBS agree.}
	\label{fig:light-all-methods}
\end{figure}

\paragraph{Concentration preservation.} We perform an analysis of the bNIRS dataset between three different algorithms for the problem of preserving the inferred concentrations (with the NRMSE criterion): a brute force computation, the SBS algorithm, and the stochastic Differential Evolution (DE) method. We first use a coarse grid for the wavelengths, having measurements only at 22 equally spaced wavelengths, and only using 10 different time points per piglet to allow for reasonable computation times. The results for all three computation optimization algorithms are shown in Fig. \ref{fig:light-all-methods}. 

We found that the stochastic DE approach is able to outperform the SBS method when the number of desired wavelengths $d$ is about one-third (or less) than the number of all measured wavelengths $n$ (see Fig. \ref{fig:light-all-methods} (left) when $d=6$). The SBS method however outperformed the DE approach for large number of desired wavelengths $d$, in our experiments when $d>9$. 
This finding is similar to what has been observed in literature with genetic algorithms \cite{Jain1997,Ferri1994}: GAs are known to fail on high dimensional problems, whereas sequential methods (such as the SBS, or SFS) are able to maintain good performance even with the curse of dimensionality. However, the SBS itself cannot be expected to perform well with a small number of desired wavelengths $d$, as iterative removal of wavelengths assumes the previous iteration to be the optimal set of wavelengths. This assumption is bound to increase the resulting error with each iteration.

We also give a brief comparison of the three optimal wavelengths found by all approaches in Fig. \ref{fig:light-all-methods} (right). Our findings are very much in line with the results from \cite{Arifler2015}, as the stochastic approach (GA in \cite{Arifler2015}, and DE in this work) obtains the same results as the brute force method for small number of wavelengths. Perhaps unsurprisingly, the optimal three wavelengths found through the brute force and DE method are also near the peaks of the three model-assumed chromophores, very similarly to the results of \cite{Arifler2015}. The SBS method finds similar wavelengths as the other two methods, with only the center wavelength differing from the oxCCO peak.

Having compared the different algorithms in a coarse setting (to allow for full brute force computation across any $d$-cardinality of wavelength sets), we now test the concentration and spectral fit preservation approaches at a higher resolution (namely measurements at 37 total wavelengths, and 40 timepoints per piglet), assuming $d \leq 10$ to allow for brute-force computation within reasonable time. 

\begin{figure}[H]
	\includegraphics[width=\textwidth]{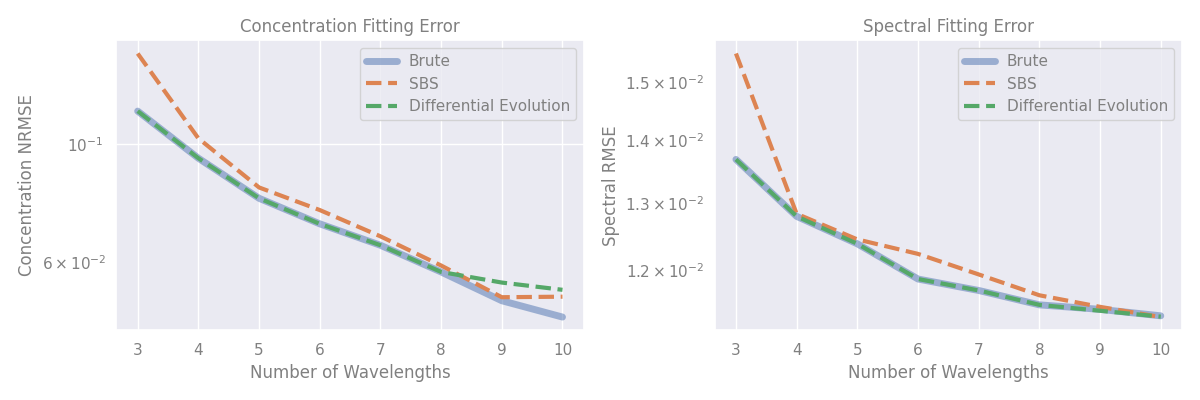}
	\caption{Development of concentration NRMSE (left) and spectral attenuation RMSE (right) fitting error, when optimizing for the concentration NRMSE in a more fine-grained scenario (i.e. using higher spectral and temporal resolution) where $d\in[3,10]$, across the three main used algorithms (brute-force search, SBS, DE).}
	\label{fig:concentration-spectral-copti}
\end{figure}

Firstly, we again analyze the performance of optimizing for the concentration NRMSE. In Fig. \ref{fig:concentration-spectral-copti}, we report the results of the observed NRMSE and spectral fitting RMSE across the three algorithms.

The error curve in Fig. \ref{fig:concentration-spectral-copti} (left) shows a similar trend to what we had observed previously. Note however that the spectral fitting error develops differently to the NRMSE, due to the previously discussed difference in linear transformation. 
The found wavelengths with the brute force method across different cardinalities $d$ are reported in Table \ref{tab:found-wl-combined} (see 'Concentration Preservation' rows). The results are presented in the same format as Table 1 from \cite{Arifler2015}, where we noticed very high similarity between the found wavelengths, as already hinted in the previous experiments.

\begin{figure}[H]
	\includegraphics[width=0.5\textwidth]{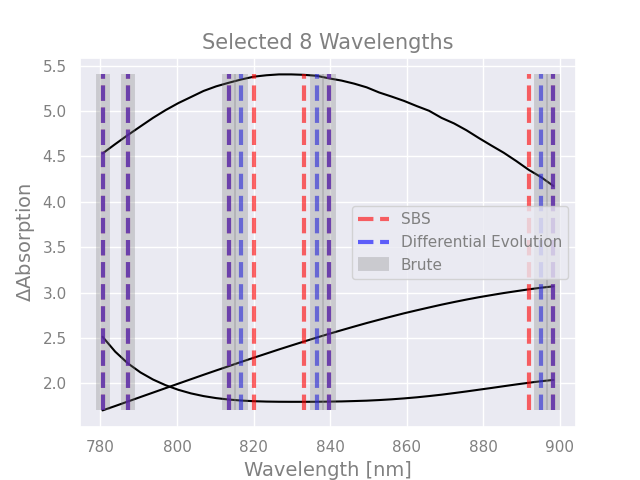}
	\includegraphics[width=0.5\textwidth]{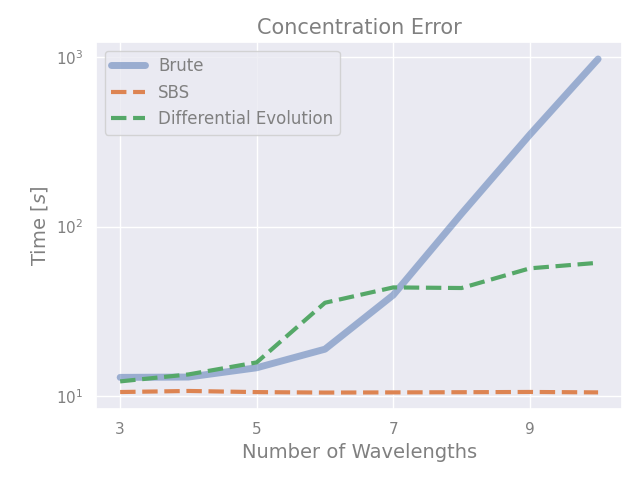}
	\caption{Selected 8 wavelengths (across the three main selection algorithms) with comparison to absorption properties of the system (left), and comparison of computational time for different cardinalities $d$ across the selection algorithms (right).}
	\label{fig:8wl-time-copti}
\end{figure}

These found wavelengths by the brute-force method are close to the peaks of the absorption of chromophores, shown in an additional example in Fig. \ref{fig:8wl-time-copti} (left) and as previously reported in \cite{Arifler2015}. For this fine-grained scenario, we also report the computation time of the three algorithms in Fig. \ref{fig:8wl-time-copti} (right).

The SBS method has a substantially lower run-time than the other two computational approaches. Since the method removes wavelengths iteratively, the highest computational time is at the beginning of the algorithm when many wavelengths need to be removed. For small wavelength number, the approach has computational time in the millisecond-second range. The brute force method grows exponentially for larger $d$. The DE method is generally faster than the brute-force method, but significantly slower than the SBS algorithm. We interestingly note that the DE method can be slower than the brute force method, as seen especially for $d\in[5,7]$. The DE algorithm, compared to the brute force method or the SBS algorithm, does not exploit the combinatorial structure of the problem. As such, the parameter search space will be much larger. Apparently, in certain cases, this can lead to longer computational times than the brute force method. However, the DE algorithm strongly outperformed the brute force approach for $d \geq 7$, as it only requires to evaluate a fraction of solutions instead of all combinations.

\begin{figure}[H]
	\includegraphics[width=\textwidth]{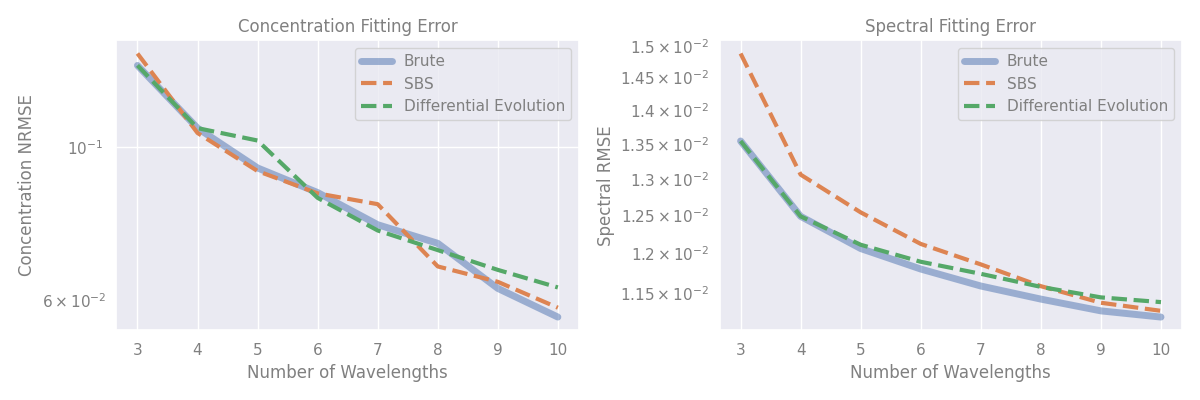}
	\caption{Development of concentration NRMSE (left) and RMSE attenuation fitting error (right), when optimizing for the spectral fit RMSE in the presented fine-grained scenario (i.e. using higher spectral and temporal resolution).}
	\label{fig:concentration-spectral-sopti}
\end{figure}

\paragraph{Spectral Fit preservation.} We now analyze the performance of optimizing for spectral fit. The concentration NRMSE and spectral fit RMSE is reported in Fig. \ref{fig:concentration-spectral-sopti}, presented analogously to Fig. \ref{fig:concentration-spectral-copti}.

\begin{figure}[H]
	\includegraphics[width=0.5\textwidth]{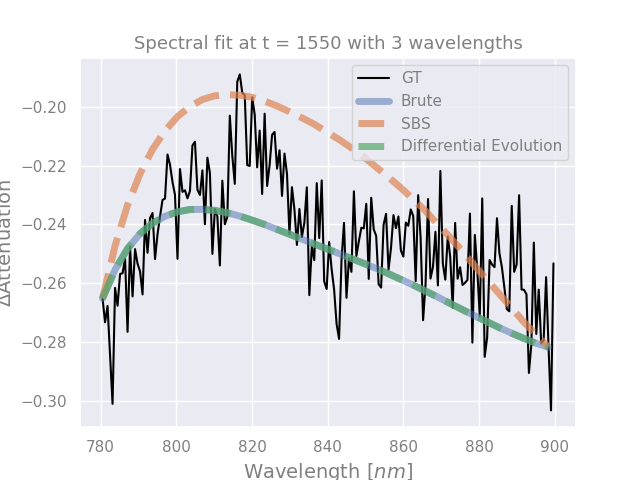}
	\includegraphics[width=0.5\textwidth]{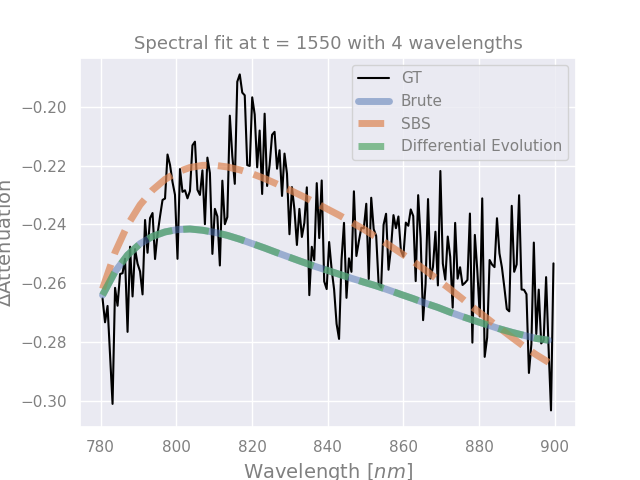}
	\caption{Comparison of the BLL-inferred attenuations, using the concentration NRMSE as optimization criterion, via the three selection algorithms using 3 wavelengths (left) and 4 wavelengths (right).}
	\label{fig:1550}
\end{figure}

We notice a smooth monotonically decreasing spectral RMSE and a less predictable behavior for the concentration NRMSE, being the opposite of what we had seen in Fig. \ref{fig:concentration-spectral-copti}. Importantly, if we observe the magnitudes of both errors in Fig. \ref{fig:concentration-spectral-copti} and Fig. \ref{fig:concentration-spectral-sopti}, we notice that they are quite similar.
This is despite the fact that 
different optimization criteria are used (preserving concentrations vs spectral fitting). Interestingly, we noticed that the SBS method performs similarly to the brute force and DE method with respect to the concentration NRMSE, especially for small number of wavelengths $d$ (see Fig. \ref{fig:concentration-spectral-sopti} (left)). 

\begin{figure}[H]
	\includegraphics[width=0.5\textwidth]{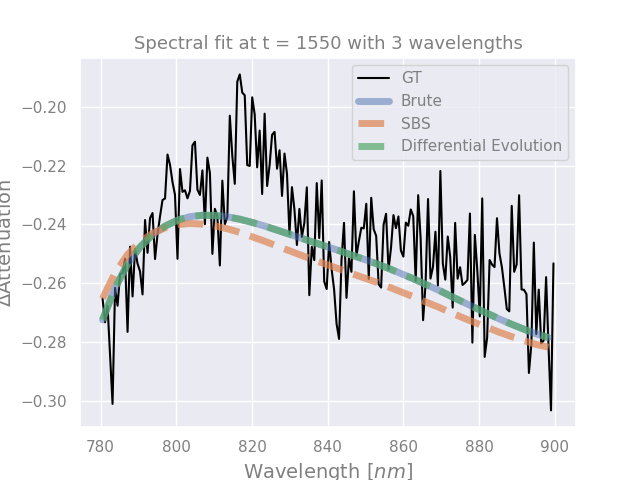}
	\includegraphics[width=0.5\textwidth]{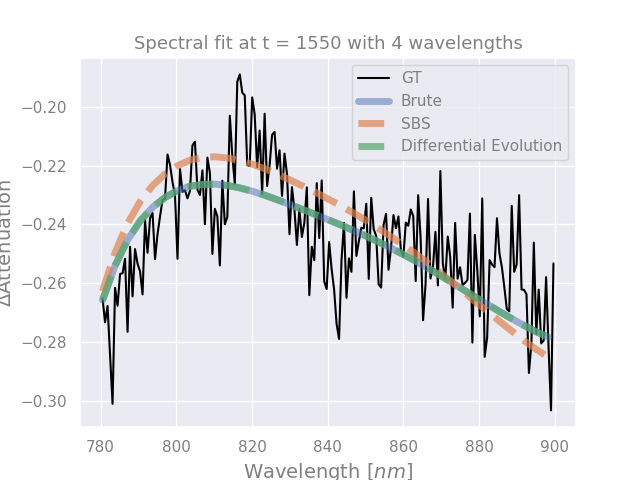}
	\caption{Comparison of the BLL-inferred attenuations, using the attenuation spectral fit as optimization criterion, via the three selection algorithms using 3 wavelengths (left) and 4 wavelengths (right).}
	\label{fig:1550s}
\end{figure}

Also, we qualitatively observed that the SBS can be on par to other methods in terms of spectral fit, when the fit is the optimization objective, see Fig. \ref{fig:1550}. When the concentration fit is the objective, the SBS can result in even more accurate fitting, as shown in Fig. \ref{fig:1550s}.


Selected wavelengths by optimizing for the spectral fit RMSE are again reported in Table \ref{tab:found-wl-combined} (see 'Spectral Fit Preservation' rows). When compared to the wavelength sets reported for 'Concentration Preservation', at $d=3$, we notice very small differences. Differences between the found sets became larger with increasing $d$, however with an absolute maximal distance of not greater than $3.4nm$.


\subsection{Model \& Data Orthogonality-based Method}
Next, we analyzed the model and data orthogonality approaches. The found wavelengths of the model orthogonality approach using the SBS method, solely removing wavelengths based on the product of singular values from the absorption matrix $M$, are again reported in Table \ref{tab:found-wl-combined} (see 'Model Orthogonality' rows). We notice high similarity, especially for $d=3$, with previously computed optimal wavelength sets from other methods, such as 'Concentration Preservation' and 'Spectral Fit Preservation'.

\begin{figure}[h]
	\includegraphics[width=\textwidth]{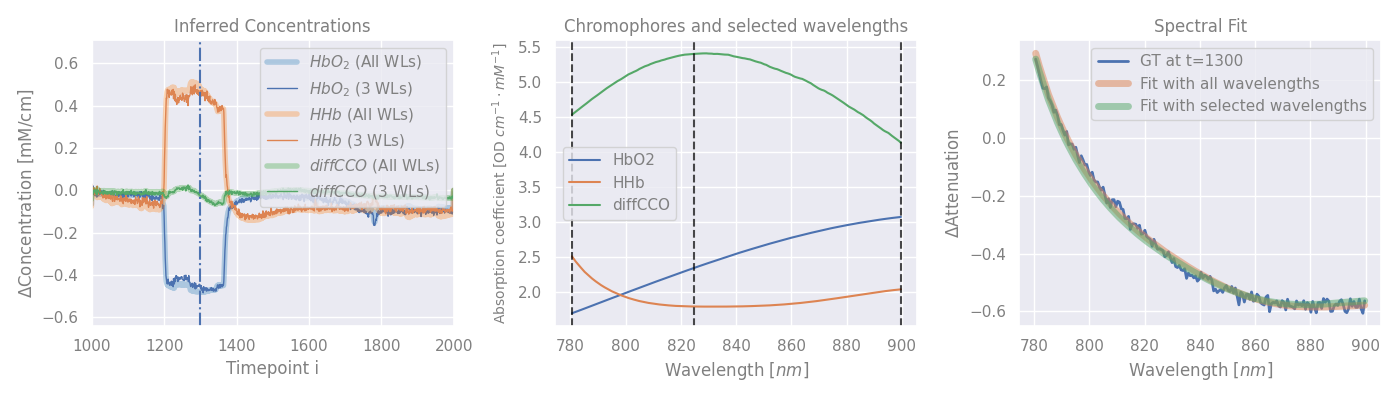}
	\caption{Performance of wavelength selection using model orthogonality. The first plot shows the inferred concentrations by using all wavelengths (All WLs) compared to inferred concentrations using the 3 found optimal wavelengths (3 WLs). The second plot shows that the found wavelengths correspond to the peaks of the three chromophores used in the model. The third plot shows the spectral fit for one exemplary timepoint, comparing the BLL-inferred attenuation using all and the three selected wavelengths with the ground truth signal.}
	\label{fig:model-ortho}
\end{figure}

The case of $d=3$ is inspected closer in Fig. \ref{fig:model-ortho}, where we compared the inferred concentrations by using all available wavelengths with this set of $3$ wavelengths. Additionally, an exemplary spectral fit is shown in Fig. \ref{fig:model-ortho}. The three wavelengths are close to the three peaks of the chromophores, which was within expectations as the wavelengths are chosen solely based on $M$. This set of wavelengths is also able to match the original inferred concentrations and spectral fits well. 

\begin{figure}[h]
	\includegraphics[width=\textwidth]{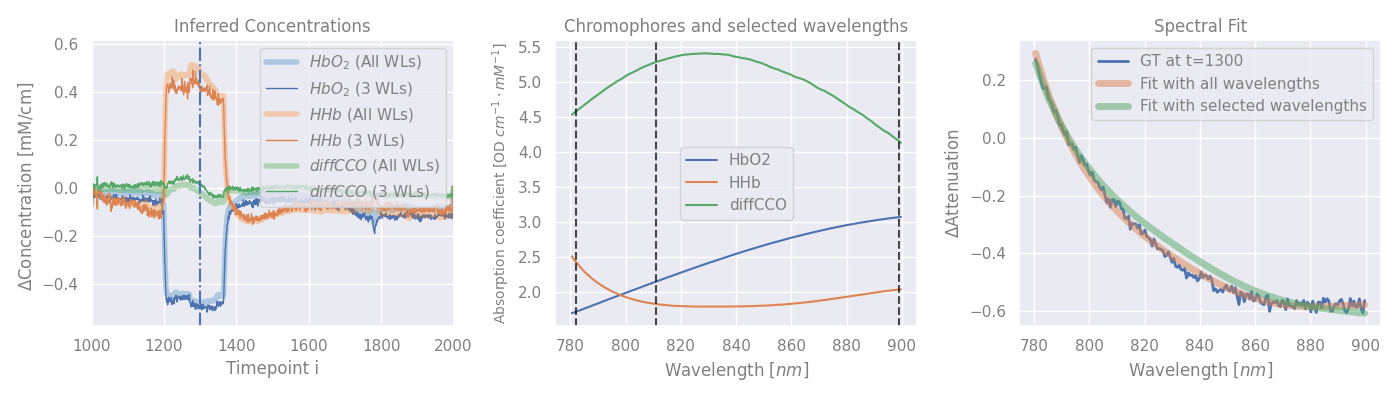}
	\caption{Performance of wavelength selection using data orthogonality. The first plot shows the inferred concentrations by using all wavelengths (All WLs) compared to inferred concentrations using the 3 found optimal wavelengths (3 WLs). The second plot shows that the found wavelengths are near the peaks of the three chromophores used in the model. The third plot shows the spectral fit for one exemplary timepoint, comparing the BLL-inferred attenuation using all and the three selected wavelengths with the ground truth signal.}
	\label{fig:data-orthogonality-780}
\end{figure}

To test the data-orthogonality method developed in section \ref{sec:a-unified-perspective}, the SBS algorithm is used solely on the measured attenuations $\Delta A$. We used 100 timepoints per piglet in this scenario, which resulted in a compute time of around 8 minutes (running in parallel on two AMD EPYC 7452 32-Core processors). The found wavelengths are again reported in Table \ref{tab:found-wl-combined} (see 'Data Orthogonality' rows), and inferred concentrations and spectral fits are shown in Fig. \ref{fig:data-orthogonality-780}.

The data-orthogonality found wavelengths qualitatively lead to some performance degradation in spectral fit and inferred concentration, which is expected as they have no information about the used model that performs the fitting. However, we found it surprising that the most informative wavelengths of this method are arguably close to all previously seen methods. We speculate that this approach suffers at finding the optimal wavelength at $825nm$, corresponding to the diffCCO peak, as CCO concentrations are considerably lower than hemoglobin. This necessarily leads to lower sensitivity to such low-concentration chromophores, compared e.g. to optimization with the NRMSE. 


\subsection{Projection Maximization for Band Selection}
We finally present the projection maximization algorithm for the band selection.
Assuming the set of chromophores of interest for given data is known, such an algorithm can be useful in finding bands which the model can describe. 

\begin{figure}[H]
	\centering
	\includegraphics[width=0.3\textwidth]{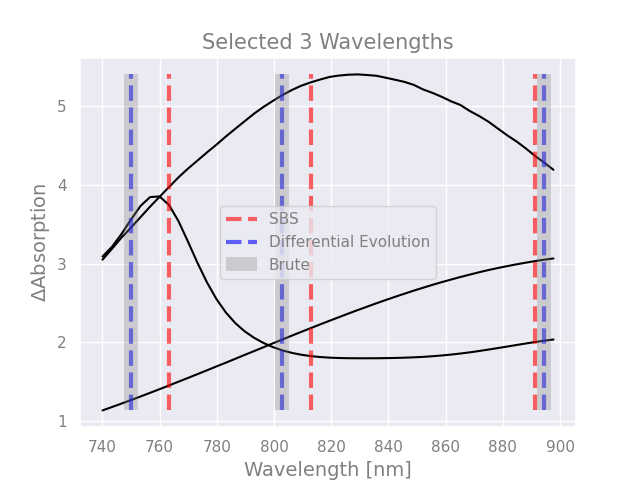}
	\includegraphics[width=0.3\textwidth]{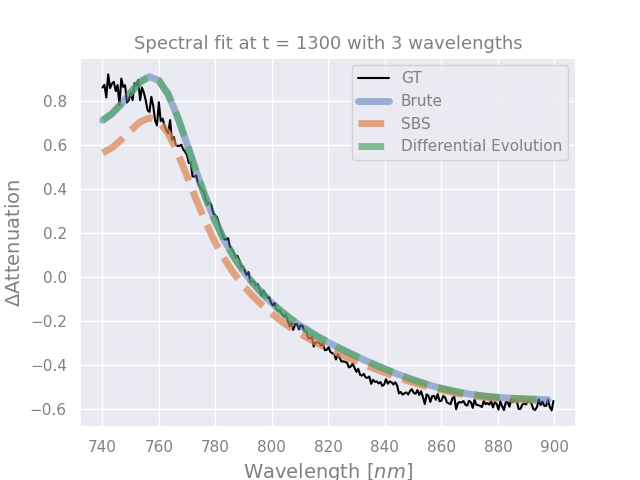}
	\includegraphics[width=0.3\textwidth]{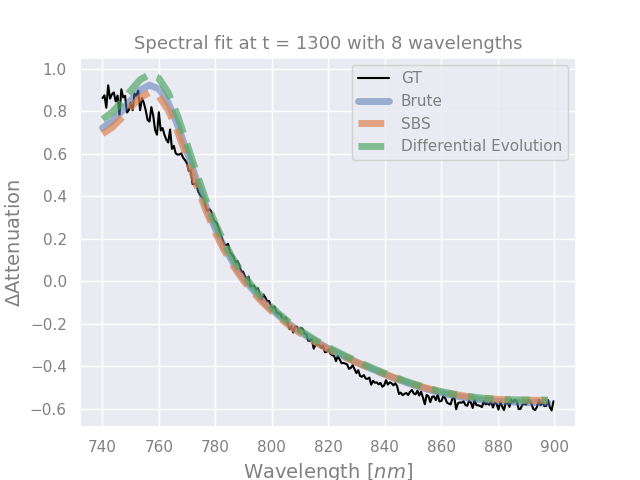}
	\caption{Selected wavelengths for the linear absorption model by NRMSE optimization in the $[740nm,900nm]$ range are not within the $[780nm,900nm]$ range (see first plot), which must deteriorate inferred concentrations due to mismatch in spectral fit between the BLL-inferred and GT attenuation below $780nm$, shown for different cardinalities $d$ in the second and third plot.}
	\label{fig:projection-nrmse}
\end{figure}

\begin{figure}[H]
	\includegraphics[width=\textwidth]{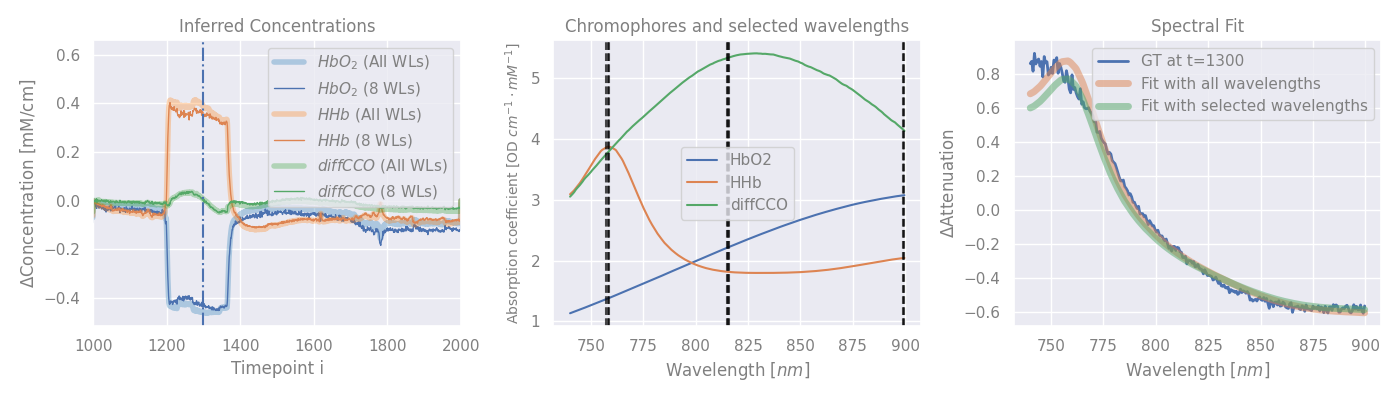}
	\includegraphics[width=\textwidth]{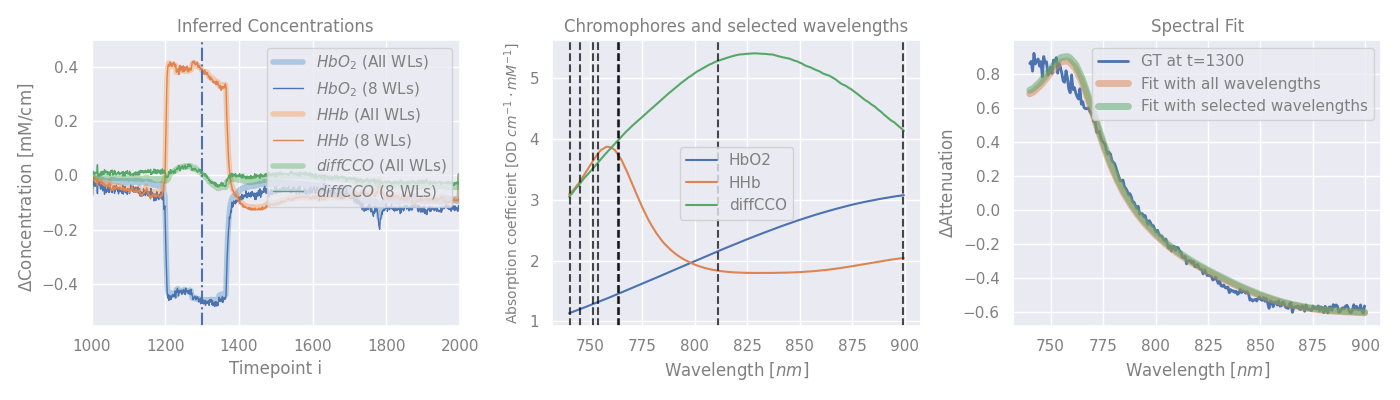}
	\caption{Performance of wavelength selection with $d=8$ using model orthogonality (top) and data orthogonality (bottom) in the $[740nm,900nm]$ range, following the same figure structure as Fig. \ref{fig:model-ortho} and Fig. \ref{fig:data-orthogonality-780}.}
	\label{fig:projection-model-data}
\end{figure}

Using measurements from the $[740nm,900nm]$ range, we analyzed the found wavelengths by the previously explored methods (NRMSE optimization, model-orthogonality, and data-orthogonality). Note that the model cannot describe the $[740nm,900nm]$ range well due to more pronounced scattering effects (which are not considered by the standard linear absorption model we use here).
The NRMSE optimization selects wavelengths below $780nm$, Fig. \ref{fig:projection-nrmse}, similarly to the model-orthogonality and data-orthogonality methods, Fig. \ref{fig:projection-model-data}.

Even though we observe a higher overlap between the found wavelengths across the conjunct methods, the chosen wavelengths fail to fit the observed signal well in the short wavelength range. We argue that these methods are not suitable for selecting wavelengths when there are physical effects not contained in the model $M$, which in practical scenarios might oftentimes be the case. Besides having unaccounted scattering effects, we could also imagine certain bands being dominated by a low signal to noise. A noise-free model would result in incorrect wavelength identification. In general, a correct spectral fit is paramount in obtaining sensible concentrations in any spectroscopic application. Both in Figure \ref{fig:projection-nrmse} and Figure \ref{fig:projection-model-data}, we qualitatively notice the spectral fit is unsatisfactory. Further, the diffCCO concentrations take positive values at the beginning of HI, which is unexpected as a monotonic decrease in diffCCO, compared to baseline, was observed in previous studies \cite{Arifler2015, Kaynezhad2019}. 

\begin{figure}[H]
	\centering
	\includegraphics[width=0.49\textwidth]{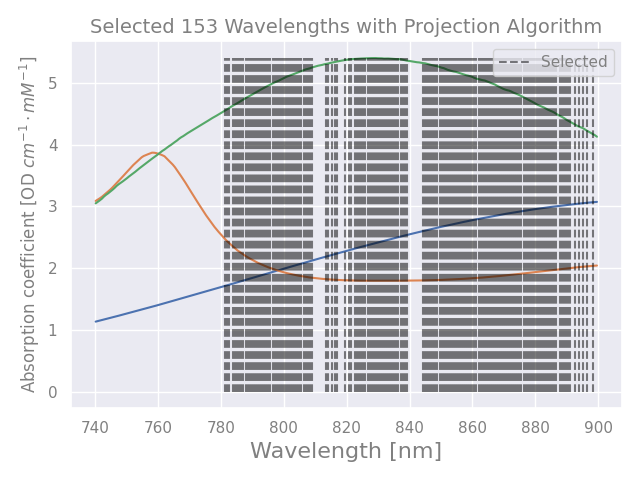}
	\includegraphics[width=0.49\textwidth]{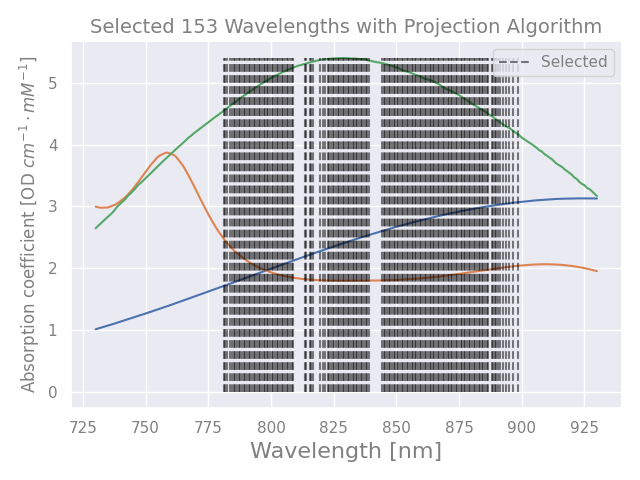}
	\caption{Selected wavelengths by the projection algorithm when selecting for $d=153$ for two different ranges: $[740nm,900nm]$ (left) and $[730nm,930nm]$ (right). The iterative removal starts with 244 and 306 acquired wavelengths for each range respectively, i.e. $d$ is halving the number of wavelengths of the second range.}
	\label{fig:projection}
\end{figure}

We now compare the performance of selecting a smaller subset of wavelengths in the $[740nm,900nm]$ range with the projection algorithm. Obtained results for the two cases are shown in Fig. \ref{fig:projection}. When using the projection algorithm in the $[740nm,900nm]$ range, all wavelengths between $[740nm,780nm]$ were removed. The method also performed well when further extending the range to $[730nm,930nm]$, excluding particularly wavelengths between $[730nm,780nm]$ and $[900nm,930nm]$. These findings should not be surprising since we had observed mismatching spectral fits in the $[740nm,780nm]$ range. While this algorithm requires knowledge of model chromophores, it allows to automatically select fitting bands, which is paramount in correctly inferring changes of molecular concentrations. 

\section{Conclusion}
In this work, we have presented an analysis of wavelength selection for optical monitoring using various methodologies, focusing on the extraction of chromophore concentration from living tissues as an application. We have shown that traditional methods, which preserve chromophore concentrations, yield similar optimal wavelengths to other methods reported in the literature (such as spectral fit preservation, or model orthogonality methods) when tested on our well-studied broadband spectroscopy dataset of piglets undergoing hypoxia-ischemia. While concentration preservation is the standard wavelength selection approach, we believe that it is important to ensure this alignment of selected wavelengths with other methods, such as spectral preservation (without a correct spectral fit to the raw data, concentration optimization alone can yield physically impossible biological estimations). We further show that purely data-driven wavelength selection methods, such as data orthogonality, can overlook important bands due to magnitude differences between chromophores. Because these methods optimize for high spectral variance, the subtle signals of low-concentration chromophores like cytochrome can be eclipsed by noise, causing the algorithm to erroneously prioritize highly noisy bands. However, in regions with a high signal-to-noise ratio, the alignment between model-based and data-driven wavelength selection methods can serve as an indicator of a biophysical model's suitability for the given spectral data.

We finally show that a novel approach, termed 'projection maximization', bridges the divide between model-driven and data-driven wavelength selection methods. This method identifies relevant wavelengths by projecting measured attenuations onto a selected model, yielding optimal wavelengths in a given dataset that satisfy model-driven constraints.

Overall, we believe that such a unified view could yield a potentially robust methodology for wavelength selection. One could first run projection maximization (or similar techniques for pre-identification of the optimal model-data matching subsets) to make sure that the biophysical model has enough capacity to describe real spectra. Subsequently, wavelength selection algorithms can be applied across the bands selected by the projection maximization to estimate a more reliable minimal wavelength subset.


\begin{backmatter}


\bmsection{Disclosures}
The authors declare no conflicts of interest.

\bmsection{Acknowledgment}
This work was supported by the European Union’s Horizon Research programme (Grant No. 101071040). This work was supported by the Gemeinnützige Hertie Stiftung.

\end{backmatter}



\bibliography{bibliography}

\begin{thebibliography}{10}
\newcommand{\enquote}[1]{``#1''}

\bibitem{Cheong1990}
W.~Cheong, S.~Prahl, and A.~Welch, \enquote{A review of the optical properties
  of biological tissues,} {\protect\JournalTitle{IEEE Journal of Quantum
  Electronics}} \textbf{26}, 2166--2185 (1990).

\bibitem{Wilson1990}
B.~Wilson and S.~Jacques, \enquote{Optical reflectance and transmittance of
  tissues: principles and applications,} {\protect\JournalTitle{IEEE Journal of
  Quantum Electronics}} \textbf{26}, 2186--2199 (1990).

\bibitem{Oshina2021}
I.~Oshina and J.~Spigulis, \enquote{Beer-lambert law for optical tissue
  diagnostics: current state of the art and the main limitations,}
  {\protect\JournalTitle{Journal of Biomedical Optics}} \textbf{26}, 100901
  (2021).

\bibitem{Pichette2016}
J.~Pichette, A.~Laurence, L.~Angulo, \emph{et~al.}, \enquote{{Intraoperative
  video-rate hemodynamic response assessment in human cortex using snapshot
  hyperspectral optical imaging},} {\protect\JournalTitle{Neurophotonics}}
  \textbf{3}, 045003 (2016).

\bibitem{Kohl2000}
M.~Kohl, U.~Lindauer, G.~Royl, \emph{et~al.}, \enquote{Physical model for the
  spectroscopic analysis of cortical intrinsic optical signals,}
  {\protect\JournalTitle{Physics in Medicine \& Biology}} \textbf{45}, 3749
  (2000).

\bibitem{Delpy1988}
D.~T. Delpy, M.~Cope, P.~van~der Zee, \emph{et~al.}, \enquote{Estimation of
  optical pathlength through tissue from direct time of flight measurement.}
  {\protect\JournalTitle{Physics in medicine and biology}} \textbf{33},
  1433--42 (1988).

\bibitem{giannoniphd}
L.~Giannoni, \enquote{Hyperspectral imaging of the haemodynamic and metabolic
  states of the exposed cortex,} Ph.D. thesis, University College London
  (2020).

\bibitem{Scholkmann2014}
F.~Scholkmann, S.~Kleiser, A.~J. Metz, \emph{et~al.}, \enquote{A review on
  continuous wave functional near-infrared spectroscopy and imaging
  instrumentation and methodology,} {\protect\JournalTitle{NeuroImage}}
  \textbf{85}, 6--27 (2014). Celebrating 20 Years of Functional Near Infrared
  Spectroscopy (fNIRS).

\bibitem{Kocsis2006}
L.~Kocsis, P.~Herman, and A.~Eke, \enquote{The modified beer-lambert law
  revisited.} {\protect\JournalTitle{Physics in medicine and biology}}
  \textbf{51}, N91--8 (2006).

\bibitem{Giannoni2018}
L.~Giannoni, F.~Lange, and I.~Tachtsidis, \enquote{Hyperspectral imaging
  solutions for brain tissue metabolic and hemodynamic monitoring: past,
  current and future developments.} {\protect\JournalTitle{Journal of optics
  (2010)}} \textbf{20}, 044009 (2018).

\bibitem{giannoni2026hyperspectral}
L.~Giannoni, F.~Lange, and I.~Tachtsidis, \enquote{Hyperspectral imaging
  solutions for brain tissue metabolic and haemodynamic monitoring: an updated
  perspective,} {\protect\JournalTitle{Journal of Optics}}  (2026).

\bibitem{Penrose1956}
R.~Penrose, \enquote{On best approximate solutions of linear matrix equations,}
  {\protect\JournalTitle{Mathematical Proceedings of the Cambridge
  Philosophical Society}} \textbf{52}, 17–19 (1956).

\bibitem{Ayala2022}
L.~Ayala, F.~Isensee, S.~J. Wirkert, \emph{et~al.}, \enquote{Band selection for
  oxygenation estimation with multispectral/hyperspectral imaging.}
  {\protect\JournalTitle{Biomedical optics express}} \textbf{13}, 1224--1242
  (2022).

\bibitem{Jain1997}
A.~Jain and D.~Zongker, \enquote{Feature selection: evaluation, application,
  and small sample performance,} {\protect\JournalTitle{IEEE Transactions on
  Pattern Analysis and Machine Intelligence}} \textbf{19}, 153--158 (1997).

\bibitem{Kaynezhad2019}
P.~Kaynezhad, S.~Mitra, G.~Bale, \emph{et~al.}, \enquote{{Quantification of the
  severity of hypoxic-ischemic brain injury in a neonatal preclinical model
  using measurements of cytochrome-c-oxidase from a miniature
  broadband-near-infrared spectroscopy system},}
  {\protect\JournalTitle{Neurophotonics}} \textbf{6}, 045009 (2019).

\bibitem{Katoch2021}
S.~Katoch, S.~S. Chauhan, and V.~Kumar, \enquote{A review on genetic algorithm:
  past, present, and future,} {\protect\JournalTitle{Multimedia Tools and
  Applications}} \textbf{80}, 8091--8126 (2021).

\bibitem{Arifler2015}
D.~Arifler, T.~Zhu, S.~Madaan, and I.~Tachtsidis, \enquote{Optimal wavelength
  combinations for near-infrared spectroscopic monitoring of changes in brain
  tissue hemoglobin and cytochrome c oxidase concentrations.}
  {\protect\JournalTitle{Biomedical optics express}} \textbf{6}, 933--47
  (2015).

\bibitem{Chauvin2022}
J.~Chauvin, A.~Akhbardeh, R.~Brunnemer, \emph{et~al.}, \enquote{Simulated
  annealing-based wavelength selection for robust tissue oxygenation estimation
  powered by the extended modified lambert-beer law,}
  {\protect\JournalTitle{Applied Sciences}} \textbf{12} (2022).

\bibitem{Kirkpatrick1983}
S.~Kirkpatrick, C.~D. Gelatt, and M.~P. Vecchi, \enquote{Optimization by
  simulated annealing,} {\protect\JournalTitle{Science}} \textbf{220}, 671--680
  (1983).

\bibitem{Martinez2019}
B.~Martinez, R.~Leon, H.~Fabelo, \emph{et~al.}, \enquote{Most relevant spectral
  bands identification for brain cancer detection using hyperspectral imaging.}
  {\protect\JournalTitle{Sensors (Basel, Switzerland)}} \textbf{19} (2019).

\bibitem{Perez2007}
R.~Perez and K.~Behdinan, \enquote{Particle swarm approach for structural
  design optimization,} {\protect\JournalTitle{Computers \& Structures}}
  \textbf{85}, 1579--1588 (2007).

\bibitem{Sharma2018}
S.~Sharma and K.~M. Buddhiraju, \enquote{Spatial–spectral ant colony
  optimization for hyperspectral image classification,}
  {\protect\JournalTitle{International Journal of Remote Sensing}} \textbf{39},
  2702--2717 (2018).

\bibitem{An2015}
X.~An, A.~W. Caswell, J.~J. Lipor, and S.~T. Sanders, \enquote{Optimized
  wavelength selection for molecular absorption thermometry.}
  {\protect\JournalTitle{Applied spectroscopy}} \textbf{69}, 464--72 (2015).

\bibitem{Storn1997}
R.~Storn and K.~Price, \enquote{Differential evolution - a simple and efficient
  heuristic for global optimization over continuous spaces,}
  {\protect\JournalTitle{Journal of Global Optimization}} \textbf{11}, 341--359
  (1997).

\bibitem{Guyon2003}
I.~Guyon and A.~Elisseeff, \enquote{An introduction to variable and feature
  selection,} {\protect\JournalTitle{J. Mach. Learn. Res.}} \textbf{3},
  1157–1182 (2003).

\bibitem{Bale2016}
G.~Bale, C.~E. Elwell, and I.~Tachtsidis, \enquote{From jöbsis to the present
  day: a review of clinical near-infrared spectroscopy measurements of cerebral
  cytochrome-c-oxidase.} {\protect\JournalTitle{Journal of biomedical optics}}
  \textbf{21}, 091307 (2016).

\bibitem{Luke2013}
G.~P. Luke, S.~Y. Nam, and S.~Y. Emelianov, \enquote{Optical wavelength
  selection for improved spectroscopic photoacoustic imaging,}
  {\protect\JournalTitle{Photoacoustics}} \textbf{1}, 36--42 (2013).

\bibitem{Marois2018}
M.~Marois, S.~L. Jacques, and K.~D. Paulsen, \enquote{{Optimal wavelength
  selection for optical spectroscopy of hemoglobin and water within a simulated
  light-scattering tissue},} {\protect\JournalTitle{Journal of Biomedical
  Optics}} \textbf{23}, 071202 (2018).

\bibitem{Eames2008}
M.~E. Eames, J.~Wang, B.~W. Pogue, and H.~Dehghani, \enquote{Wavelength band
  optimization in spectral near-infrared optical tomography improves accuracy
  while reducing data acquisition and computational burden,}
  {\protect\JournalTitle{Journal of Biomedical Optics}} \textbf{13}, 054037
  (2008).

\bibitem{Sun2022}
J.~Sun, Z.~Wu, L.~Wang, \emph{et~al.}, \enquote{Band selection for mapping
  chromophores of skin tissue,} {\protect\JournalTitle{J. Biophotonics}}
  \textbf{15}, e202200038 (2022).

\bibitem{Corlu2003}
A.~Corlu, T.~Durduran, R.~Choe, \emph{et~al.}, \enquote{Uniqueness and
  wavelength optimization in continuous-wave multispectral diffuse optical
  tomography,} {\protect\JournalTitle{Opt. Lett.}} \textbf{28}, 2339--2341
  (2003).

\bibitem{Corlu2005}
A.~Corlu, R.~Choe, T.~Durduran, \emph{et~al.}, \enquote{Diffuse optical
  tomography with spectral constraints and wavelength optimization,}
  {\protect\JournalTitle{Appl. Opt.}} \textbf{44}, 2082--2093 (2005).

\bibitem{Correia2010}
T.~Correia, A.~Gibson, and J.~Hebden, \enquote{Identification of the optimal
  wavelengths for optical topography: a photon measurement density function
  analysis.} {\protect\JournalTitle{Journal of biomedical optics}} \textbf{15},
  056002 (2010).

\bibitem{Arridge1998}
S.~R. Arridge and W.~R.~B. Lionheart, \enquote{Nonuniqueness in diffusion-based
  optical tomography,} {\protect\JournalTitle{Opt. Lett.}} \textbf{23},
  882--884 (1998).

\bibitem{Mazhar2010}
A.~Mazhar, S.~Dell, D.~J. Cuccia, \emph{et~al.}, \enquote{Wavelength
  optimization for rapid chromophore mapping using spatial frequency domain
  imaging,} {\protect\JournalTitle{Journal of Biomedical Optics}} \textbf{15},
  061716 (2010).

\bibitem{Luke2014}
G.~P. Luke and S.~Y. Emelianov, \enquote{Optimization of in vivo spectroscopic
  photoacoustic imaging by smart optical wavelength selection.}
  {\protect\JournalTitle{Optics letters}} \textbf{39}, 2214--7 (2014).

\bibitem{Chang2006}
C.-I. Chang and S.~Wang, \enquote{Constrained band selection for hyperspectral
  imagery,} {\protect\JournalTitle{IEEE Transactions on Geoscience and Remote
  Sensing}} \textbf{44}, 1575--1585 (2006).

\bibitem{Nouri2016}
D.~Nouri, Y.~Lucas, and S.~Treuillet, \enquote{Hyperspectral interventional
  imaging for enhanced tissue visualization and discrimination combining band
  selection methods,} {\protect\JournalTitle{International Journal of Computer
  Assisted Radiology and Surgery}} \textbf{11}, 2185--2197 (2016).

\bibitem{Peng2005}
H.~Peng, F.~Long, and C.~Ding, \enquote{Feature selection based on mutual
  information criteria of max-dependency, max-relevance, and min-redundancy,}
  {\protect\JournalTitle{IEEE Transactions on Pattern Analysis and Machine
  Intelligence}} \textbf{27}, 1226--1238 (2005).

\bibitem{Ding2005}
C.~Ding and H.~Peng, \enquote{Minimum redundancy feature selection from
  microarray gene expression data.} {\protect\JournalTitle{Journal of
  bioinformatics and computational biology}} \textbf{3}, 185--205 (2005).

\bibitem{Bahl2023}
A.~Bahl, S.~Segaud, Y.~Xie, \emph{et~al.}, \enquote{A comparative study of
  analytical models of diffuse reflectance in homogeneous biological tissues:
  Gelatin based phantoms and monte carlo experiments,}  (2023).

\bibitem{Ezhov2023}
I.~Ezhov, L.~Giannoni, S.~Shit, \emph{et~al.}, \enquote{Identifying chromophore
  fingerprints of brain tumor tissue on hyperspectral imaging using principal
  component analysis,} in \emph{European Conference on Biomedical Optics,}
  (2023).

\bibitem{ezhov2024shallow}
I.~Ezhov, K.~Scibilia, L.~Giannoni, \emph{et~al.}, \enquote{{Learnable
  real-time inference of molecular composition from diffuse spectroscopy of
  brain tissue},} {\protect\JournalTitle{Journal of Biomedical Optics}}
  \textbf{29} (2024).

\bibitem{Ferri1994}
F.~Ferri, P.~Pudil, M.~Hatef, and J.~Kittler, \enquote{Comparative study of
  techniques for large-scale feature selection,} in \emph{Pattern Recognition
  in Practice IV,}  vol.~16 of \emph{Machine Intelligence and Pattern
  Recognition} E.~S. GELSEMA and L.~S. KANAL, eds. (North-Holland, 1994), pp.
  403--413.

\end{thebibliography}

\end{document}